\documentstyle[12pt]{article}
\begin{document}
\def\thebibliography#1{\section*{REFERENCES\markboth
 {REFERENCES}{REFERENCES}}\list
 {[\arabic{enumi}]}{\settowidth\labelwidth{[#1]}\leftmargin\labelwidth
 \advance\leftmargin\labelsep
 \usecounter{enumi}}
 \def\newblock{\hskip .11em plus .33em minus -.07em}
 \sloppy
 \sfcode`\.=1000\relax}
\let\endthebibliography=\endlist

\hoffset = -1truecm
\voffset = -2truecm


\title{\large\bf
A Dynamical Principle For The  Salpeter Equation  
}
\author{
{\normalsize \bf
A.N.Mitra \thanks{e.mail: ganmitra@nde.vsnl.net.in }
}\\
\normalsize 244 Tagore Park, Delhi-110009, India.} 

\date{}

\maketitle


\begin{abstract}

The `Salpeter Equation' which  has long been known as the 3D 
version of the 4D Bethe-Salpeter Equation under the Instantaneous 
Approximation, has a well-defined rationale  that stems from the 
half-century old Markov-Yukawa Transversality Principle $(MYTP)$ which not 
only effects an exact 3D reduction from the  original (4D) BS form, 
but also provides an equally exact $reconstruction$ of the 4D BS amplitude 
in terms of 3D ingredients. The second aspect which is $new$, opens up
a vista of applications to transition amplitudes as 4D loop integrals.  
  


\end{abstract}

\section*{1.  Introduction: Historical Background}
   
The problem of probability interpretation of the well-known Bethe-Salpeter
Equation (BSE) [1] has for long been a much discussed issue in the Physics 
literature, and has led to various attempts [2-5] towards its 3D reduction,
as well as for intrinsic 3D formulations  of various kinds [6]. On the 
other hand, the issue should be viewed against the theoretical perspective
of the BSE as only an $approximate$ description (in the so-called ladder
approximation) of the equations of motion that follow strictly from the 
standard QED Lagrangian, viz., the Schwinger-Dyson Equations (SDE) which 
are an infinite chain  of  equations connecting successively higher order 
vertex functions. Thus a conceptual problem like lack of probability
interpretation for a $truncated$ two-body system need hardly cause undue
surprise. Nevertheless, the approximate nature of the BSE [1] has never
been a detracting issue from its practical usefulness which has lent
considerable credence to the attempts at its 3D reduction designed among
other things to restore the probability interpretation to the BS amplitude.
Of these attempts, the Salpeter Equation [3] is perhaps the oldest, and
has claimed considerable attention in the contemporary literature, from
the atomic two-body problem to the  QCD context of heavy quarkonia. 
Another possible motivation for the 3D reduction of the BSE is 
the observed $O(3)$-like spectra of the respective energy levels.
\par
	To meet this dual requirement of both the probability 
interpretation as well as the observed spectra, the "Instantaneous 
Approximation"  was historically [2] the earliest ansatz to be invoked for 
a 3D reduction of the full 4D BSE, leading to the Salpeter equation [3]. 
However, the ansatz (which also suffered from lack of Lorentz Covariance) 
seemed to beg a solid underlying principle. On the other hand, the 
extensive use  of the Salpeter equation  in the contemporary literature
demands a firmer theoretical foundation. 

\subsection*{1.1 The Markov-Yukawa Transversality Principle}

In the quest for a theoretical basis for the Instantaneous Approximation, 
success seems to have come from a rather unexpected quarter, viz., the 
half-century old Markov-Yukawa Transversality Principle (MYTP) [7], as
was to be discovered by the Dubna Group [8]. Now the MYTP [7] presupposes a
dependence of the field variable on both $x$ and $p$. While this is
unacceptable for an elementary particle description (and is probably the
reason why the Yukawa non-local field thery [7b] was found unattractive),
it seems to be  ideally suited to a  $composite$ particle description, 
wherein the momentum dependence comes from the direction of the total 
4-momentum $P_\mu$.  For a bilocal field ${\cal M}(z,X)$, the 
Transversality condition [7] was shown by Lukierski et al [9] to be 
equivalent to a `gauge principle' which expresses the redundance of the
$longitudinal$ component of the relative momentum for the physical
interaction between the two constituents. As will be shown in Sect.2,
this condition amounts to a covariant 3D support to the input 4-quark 
Lagrangian, whence follows the 4D BSE with a 3D kernel support governed by
Covariant Instantaneity. This Principle was first invoked by the Dubna 
group [8] to show that the 3D Salpeter equation [3] follows as an exact 
consequence of the covariant 3D support to the Bethe-Salpeter kernel,
with the preferred direction as $P_\mu$. This gave a firm theoretical 
basis to the 3D Salpeter equation.
\par
	The other side of the coin, apparently missed by the Dubna Group
[8], concerns the question of whether the information on the 4D content
of the original BSE is retrievable, after  the 3D reduction. As was to 
be found soon afterwards [10],  the inbuilt structure of MYTP [7] ensures
that the original 4D BSE is exactly recovered by retracing the steps !  
This  two-way interconnection [10] between the 3D and 4D BSE forms was 
initially proved [10] for an idealized spinless fermion problem, but,
as will be shown in this paper, the logic goes through equally well for 
spinor fermions, thus facilitating an exact reconstruction of the 
original 4D BS amplitude in terms of the 3D ingredients of the Salpeter 
equation. This is, surprisingly enough, a $new$ result, considering the 
fact that this aspect of the Salpeter equation has never seen the light
of the day despite its half century old existence. A generalization of  
the 3D-4D interconnection of Bethe-Salpeter amplitudes under MYTP [7]
to the 3-body problem has been given recently [11].   
\par
	The paper is organized as follows. In Sect.2, starting with the 
logic of the MYTP [7] which mandates a covariant 3D support to the kernel of 
a BSE, we first recapulate the main steps [10] that lead to an exact
3D-4D interconnection between the corresponding BS amplitudes for a 
`spinless' two-particle system.  In Sect.3 we outline a corresponding
derivation for the Salpeter equation by recalling the main steps leading
from the 4D to the 3D form [3], and reversing these steps. Sect.4 
concludes with some comments on the significance of this results vis-a-vis
the Markov-Yukawa Principle, especially the applicability of the 
`Salpeter Vertex fn' to transition amplitudes as 4D loop integrals.

\section*{2.  MYTP As A Gauge Principle}

The logic of MYTP [7] may be traced to Yukawa's non-local field theory [7b], 
characterized by the field dependence on both coordinate and momentum.  
As this violates local micro-causality, this concept as a basic theory of
elementary particles did not find much favour within the physics community.
However this (limited) perspective had to change with the advent of QCD [12] 
which pushed the status of hadrons from the elementary to a composite level,
and gave rise to the concept of bilocal fields [13]. Within such a bilical 
scenario, the total 4-momentum $P_\mu$ of the composite hadron provides a
naturally preferred direction which forms the basis for a covariant 3D
support to the interaction kernel [8,10]. 
\par
	An important feature of bilocal dynamics is the redundance [9]
of the relative `time' variable $x_0$, ($x=x_1-x_2$), whose covariant
definition is just the longitudinal component of $x_\mu$ in the direction 
of $P_\mu$, viz.,  $x.P P_\mu/P^2$. This `redundance' is expressed 
by the statement that a translation of the relative coordinate [9]
$x_\mu \rightarrow x_\mu' + \xi P_\mu$ on the bilocal field ${\cal M}(x,P)$:
$${\cal M}(x_\mu,P_\mu) \rightarrow {\cal M}_\xi(x_\mu,P_\mu)
= {\cal M}(x_\mu+\xi P_\mu,P_\mu)$$,
which is a  sort of `gauge transformation' of the bilocal field [9], 
should leave this quantity $invariant$. This invariance is just 
the content of the Markov-Yukawa subsidiary condition [7] which, 
under an interchange of the relative coordinates and the momenta reads 
as [9, 8b]
\begin{equation}\label{1}
P_\mu {{\partial} \over {\partial x_\mu}} {\cal M}(x_\mu,P_\mu)=0 
\end{equation} 
where the direction $P_\mu$ guarantees an irreducible representation of the 
Poincare' group for the bilocal field ${\cal M}$ [9]. An equation of
type has been used in other approaches to bilocal field dynamics (see
ref [9] for other references), but this `gauge' interpretation of the
subsidiary condition [9] provides a more transparent view of the same
condition which we have abbreviated as MYTP above. 
\par
	Eq.(1) amounts to an effective 3D support to the interaction
between the constituents of the bilocal field, which may be alternatively
postulated directly for the pairwise BSE kernel $K$ [10] by demanding that 
it be a function of only ${\hat q}_\mu = q-q.PP_\mu/P^2$, which implies that 
${\hat q}.P \equiv 0$. In this approach, the propagators are left untouched 
in their full 4D forms.  This is somewhat complementary to the 3D BSE 
reduction methods [4-6] (propagators manipulated but kernel left untouched),
so that the resulting equations [8,10] look rather unfamiliar vis-a-vis 
3D BSE's [4-6], but it has the advantage of allowing a $simultaneous$ 
use of both 3D and 4D BSE forms via their interlinkage. Indeed what 
distinguishes the Covariant Instantaneity Ansatz [10] from the more familiar 
3D BSE reductions [4-6] is its capacity for a 2-way linkage: an exact 3D BSE 
reduction, and an equally exact reconstruction of the original 4D BSE form 
without extra charge [10].  In contrast the more familiar methods of BSE 
reduction [4-6] give at most a one-way connection, viz., 
a $4D\rightarrow 3D$ reduction, but not vice versa. This is a plus
point for MYTP, and may well have a wider significance than the mere BSE 
context above, as an effective dynamics for strong interactions. 

\subsection*{2.1  3D-4D Interconnection: Spinless Particles} 

To demonstrate the basic 3D-4D interconnection  under MYTP [7], consider  
a system of two dentical spinless particles, with the BSE [10]  
\begin{equation}\label{2}
 i(2\pi)^4 \Phi(q,P) = (\Delta_1\Delta_2)^{-1} \int d^3{\hat q}'M d\sigma' 
K({\hat q},{\hat q}') \Phi(q',P);  [\Delta_{1,2} = m_q +p_{1,2}^2]
\end{equation}
where the 3D support to the kernel $K$ is implied in its `hatted' structure:
\begin{equation}\label{3}
{\hat q}_\mu= q_\mu- \sigma P_\mu; \sigma= q.P/P^2; {\hat q}.P \equiv 0.
\end{equation}                                                                            

The relative and total 4-momenta are related by

$$p_1+p_2 = P = p_1'+p_2'; 2q = p_1-p_2; 2q'=p_i'-p_2'.$$     

The 3D wave function $\phi({\hat q})$ is defined by [10]
\begin{equation}\label{4}
 \phi({\hat q}) = \int M d\sigma \Phi(q,P) 
\end{equation}
When (4) is substituted on the RHS of (2) one gets
\begin{equation}\label{5}
 i(2\pi)^4 \Phi(q,P) = (\Delta_1\Delta_2)^{-1} \int d^3{\hat q}' 
K({\hat q},{\hat q}') \phi({\hat q}')
\end{equation}

Now integrate both sides of this equation wrt $\sigma$ to get an explicit
3D equation
\begin{equation}\label{6}
 (2\pi)^3 D({\hat q}) \phi({\hat q}) = 
\int d^3{\hat q}' K({\hat q},{\hat q}') \phi({\hat q}')           
\end{equation}
where the 3D denominator function is given by
\begin{equation}\label{7}
 2i\pi D^{-1}({\hat q}) = \int M d\sigma (\Delta_1 \Delta_2)^{-1} 
\end{equation}

A comparison of (5) with (6) via (7)  gives the 3D-4D interconnection  
\begin{equation}\label{8}
 21\pi \Delta_1 \Delta_2 \Phi (q,P) = D({\hat q}) \phi({\hat q})
\end{equation}
which directly identifies the RHS as the hadron-quark Vertex Function 
\begin{equation}\label{9}
\Gamma = D \times \phi /2i\pi.                                   
\end{equation}

\section*{3.  Salpeter Eqn: 3D-4D Interlinkage}

Let us now look at the Salpeter equation for the relativistic hydrogen 
atom problem, which in the notation of the original paper [3] reads as
 
\begin{equation}\label{10} 
i\pi^2 F(q_\mu) \psi(q) = \alpha \int d^4 k {\bf k}^{-2} \psi(q+k) 
\end{equation}          
A comparison of this equation with eq.(6) shows a precise correspondence,
except for certain technicalities arisin from its fermionic content. Indeed 
it stems from an equation of the form (2), where the 3D kernel support is 
due to the (non-covariant) instantaneous (adiabatic) assumption [3], 
manifesting from its dependence on the 3-vector ${\bf k}$, while 
the quantity $F(q_\mu)$ plays just the role of the product of the two 4D 
propagators $\Delta_1$ and $\Delta_2$ in (2):
\begin{equation}\label{11}
 F(q) = (\mu_1 E -H_1({\bf q})+\epsilon)(\mu_2 E -H_1({\bf q})-\epsilon)  
\end{equation}
with the time-like components identified as the $\epsilon$ terms !
Next, define the 3D wave function $\phi({\bf q})$ by
\begin{equation}\label{12}
\phi({\bf q}) = \int d\epsilon \psi ({\bf q}, \epsilon)
\end{equation}          
which is the counterpart of (4), and use this result to integrate both
sides of (10) wrt $\epsilon$, after dividing by $F(q)$, so as to get
the 3D Salpeter equation [3]
\begin{equation}\label{13}
 [E-H_1({\bf q})-H_2({\bf q})]\phi_{\pm \pm} = 
\pm \Lambda_{\pm}^{(1)}\Lambda_{\pm}^{(2)} (-2i\pi \Gamma({\bf q})
= (-4i\alpha) \int d^3 k {\bf k}^{_2} \phi({\bf q+k})
\end{equation}
where the $\pm$ components are associated with the energy projection 
operators $\Lambda$ which however do not involve the time-like $\epsilon$.

The new aspect, on the other hand, is the 3D-4D interconnection 
which is obtained by substituting the second part of eq.(13) 
on the RHS of (10), after making use of (12):
\begin{equation}\label{14}
 F(q)\psi(q) = \Gamma({\bf q}) 
\end{equation}
where $\Gamma({\bf q})$ is the 3D BS vertex function. It is the precise
fermionic counterpart of the scalar eq.(9), since the $F(q)$ function is the 
product of the two 4D propagators. The form (14) is not formally covariant, 
but this is a mere technicality which can be remedied by standard methods;
see e.g., ref [14]. 
\par
	The more interesting thing about this demonstration is the exciting 
prospect of using the  reconstructed  4D `Salpeter vertex function' (14)
as a basic ingredient for the calculation of various types of transition 
amplitudes as 4D  loop integrals  by standard Feynman techniques without 
having to face the usual problems of probabilty interpretation and/or 
spectroscopy, both of which  are now subsumed in the 3D equation (13).
This gives a sort of `two-tier' description, the 3D form (13) just right
for spectroscopy, energy levels, etc, while the 4D form (14) provides
the proper vehicle for 4D loop integrals. It is only the first (3D) part
of the Salpeter Equation [3] that has so far been evidenced in the
contemporary literature, but the second (4D) aspect is entirely $new$.
       
\section*{4.  Retrospect And Summary}

	In retrospect, we have attempted to project an aspect of the
well-known Salpeter equation [3], which had remained obscured from view
for decades, viz., a theoretical basis for its underlying Instantaneous
Approximation, offered by the Markov-Yukawa Transversality Principle [7]:
An in-built MYTP [7] in the 4D BSE structure leads to an exact 3D reduction
which, as first shown by the Dubna group [8], is a covariant generalization 
of the Salpeter Equation [3]. The (new) complementary aspect of MYTP is
its in-built capacity to $reconstruct$ with equal ease [10], the 4D vertex 
function, (9) or (14), in terms of 3D ingredients, which allows access
to transition amplitues of diverse types as 4D loop integrals. This offers
a two-tier description for the Salpeter Equation, analogously to the
quark-level hadronic BSE problem that has been under study for several
years, with its 3D form  providing access to spectroscopy [15], and the
4D form offering applications to processes like e.m. form factors [14], 
within a single framework. This dual feature distinguishes MYTP  from most 
other 3D approaches to strong interaction dynamics [4-6] which give at most 
a one-way connection (4D to 3D).  This remarkable property of 3D-4D 
interlinkage enjoyed by the Salpeter equation [3], by virtue of its 
compliance with MYTP, should hopefully  offer new incentives for its 
(second stage) applications to  4D loop integrals in a covariant manner.    
\par
	The 3D-4D interlinkage offered by MYTP is also generalizable to a
3-body BSE with pairwise kernels under covariant 3D support [11]. A second 
type of generalization of MYTP is to the covariant null-plane [14] which 
facilitates trouble-free evaluation of form factors with triangle loops.
\par
	To summarise, the instantaneous approximation which characterizes 
the Salpeter equation, comes as a mere consequence of the Markov-Yukawa 
Transversality Principle which by its very definition gives a 
precise 3D support to the BSE kernel. Secondly, MYTP  allows reconstruction 
of the 4D Salpeter amplitude in terms of 3D ingredients, a 
property which had remained obscured from view  so far. Thus the Salpeter 
equation is amenable to a two-tier formalism, the 3D form  for spectroscopy, 
and the reconstructed 4D vertex function for  4D loop integrals. Finally,
the non-covariance of the Salpeter equation [3] is a mere technicality 
which is easily remedied by standard techniques [14].  
\par
	A preliminary version of this work was reported at the
XVI Few-Body Conf at Taiwan.

\end{document}